# A Comparison of Trojan Virus Behavior in Linux and Windows Operating Systems

Ghossoon. M. W. Al-Saadoon

Ass. Professor, Head Dep.of MIS:
College of Administrative Science, Applied Science University
Manama, Kingdom of Bahrain

dr.ghoson@asu.edu.bh

Hilal M.Y. Al-Bayatti

Prof. of Computing, College of Arts and Science, Applied Science University
Manama, Kingdom of Bahrain

dr.Hilal@asu.edu.bh

Abstract—Trojan virus attacks pose one of the most serious threats to computer security. A Trojan horse is typically separated into two parts – a server and a client. It is the client that is cleverly disguised as significant software and positioned in peer-to-peer file sharing networks, or unauthorized download websites. The most common means of infection is through email attachments. The developer of the virus usually uses various spamming techniques in order to distribute the virus to unsuspecting users. Malware developers use chat software as another method to spread their Trojan horse viruses such as Yahoo Messenger and Skype.

The objective of this paper is to explore the network packet information and detect the behavior of Trojan attacks to monitoring operating systems such as Windows and Linux. This is accomplished by detecting and analyzing the Trojan infected packet from a network segment -which passes through email attachment- before attacking a host computer.

The results that have been obtained to detect information and to store infected packets through monitoring when using the web browser also compare the behaviors of Linux and Windows using the payload size after implementing the Wireshark sniffer packet results. Conclusions of the figures analysis from the packet captured data to analyze the control bits and , check the behavior of the control bits, and the usability of the operating systems Linux and Windows.

## I. INTRODUCTION

A Trojan horse is a program in which malicious or harmful code is contained inside apparently harmless program or data in such a way that it can get control and do its chosen form of damage, such as ruining or erasing data on the hard drive. A Trojan can cause massive harm to computing systems and worse still, may turn computing system into a killing machine.

A Trojan can cause massive harm to computer systems and worse yet, may turn a system into a killing machine as well. Let us look at Back Orifice specifically so we can highlight why a tool like this can become ugly if installed on your systems.

A Trojan virus works by hiding within a set of seemingly useful software programs. Once executed or installed in the system, this type of virus will start infecting other files in the computer. A Trojan virus is also usually capable of stealing important information from the user's computer. The developer will then be able to gain a level of control over the computer through the Trojan virus [6]. While these things are taking place, the user will notice that the infected computer has become very slow or unexpected windows pop up without any activity from the user. Later on, this will result in a computer crash [3].

## II. RELATED WORK

Internet security is an important element in networking. It needs protection against intruders. Even though many anti-virus software packets have been designed to detect malicious codes, they still fail to do so. There are two common methods that an anti-virus software application uses to detect viruses. The first, and by far the most common method of virus detection , to use a list of virus signature definitions; the second method is to use a heuristic algorithm to find viruses based on common behaviors. The use of heuristic algorithm involves inspecting the code in a file (or other object) to see if it contains virus-like instructions [10].

Back Orifice consists of two key pieces: a client application and a server application. The way in which Back

Orifice works is that the client application runs on one machine and the server application runs on a different machine. The client application connects to another machine using the server application. The confusing part is the server installed on the victim. Many people may be confused by this because it does not seem logical, but that is how it works. The only way for the server application of Back Orifice to be installed on a machine is for it to be installed deliberately. Obviously, the Trojan does not come with a default installation of Windows 2000, so you must find a way to get the victim to install it [7].

### III. TROJAN HORSE PROBLEMS

Trojans are difficult to detect because they appear to be useful programs or application and user tend to download them. Furthermore that database Trojans represents a sophisticated attack because the attack is separated into two parts: the injection of the malicious code and then calling it, - which is one of the reasons for Trojans being difficult to track.

This paper focuses on how it is possible to detect and analyze packet network segments through e-mail attachments, and gives a behavior comparison between windows and Linux operating systems against the Trojan attacks. This will be done through network packet information capture, check, analysis, store and display.

### IV. TYPES OF TROJAN

There are various types of Trojans that damage victim machines or threaten data integrity, or impair the functioning of the victim's machine. Multi-purpose Trojans are also included some virus writers have created multi-functional Trojans rather than Trojan packs. Some types of Trojans as listed below; this research focused on the Backdoor type.

- PSW Trojan [1,2].
- Trojan Droppers [8].
- Rootkits [2].
- Arcbomb [8].
- Trojan Downloaders [8].
- Trojan Proxies [8].
- Trojan Spies [8].
- Trojan Notifiers [8].
- Backdoors.

A. Backdoors Trojains are the most dangerous type of Trojan and also the most widespread one. These Trojans are remote administration utilities that open infected machines to external control via a LAN or the Internet. They function in the same way as legal remote administration programs used by system administrators. This makes them difficult to detect. The only difference between a legal administration tool and a backdoor is that backdoors are installed and launched without the knowledge or consent of the user of the victim machine.

B. Once the backdoor is launched, it monitors the local system without the user's knowledge; often the backdoor will not be visible in the log of active programs. Once a remote administration utility has been successfully installed and launched, the victim machine is wide open. Backdoor functions can include [6]:

- Sending/ receiving files

- Launching/ deleting files

- Executing files

- Displaying notification

- Deleting data

- Rebooting the machine

C. In other words, backdoors are used by virus writers to detect and download confidential information, execute malicious code, destroy data, including the machine in both networks and so forth. In short, backdoors combine the functionality of most other types of Trojans in one package. Backdoors have one especially dangerous sub-class: variants that can propagate like worms [6, 9]. The only difference is that worms are programmed to propagate constantly, whereas these 'mobile' backdoors spread only after a specific command from the 'master'.

### V. EXISTING NETWORK PACKET MONITORING TOOL ON GNU/LINUX

Most commonly used desktop based network monitoring tools are Tcpdump and Wireshark [10], the main features of the Wireshark are:

1. It is distributed under the Gnu's Not UNIX (GNU) General Public License (GPL) Open-source license.
2. It works in promiscuous and non-promiscuous modes.
3. It can capture data from the network or read from a capture file.
4. It has an easy-to-read and configurable GUI.
5. It has rich display filter capabilities.
6. It runs on over 20 platforms, including Uniplexed Information and Computing.
7. System (UNIX)-based operating systems (OSs), Windows, and there are third-party packages available for Mac OS X.
8. It supports over 750 protocols, because it is open source, new ones are contributed frequently.
9. It can capture data from a variety of media (e.g., Ethernet, Token-Ring, 802.11 Wireless, and so on).

10. It includes a command-line version of the network analyzer called *tshark*.

## VI. THE PROPOSED SOLUTION

The aims of this paper as mentioned before are to capture computer network packets from a network segment, check each packet for Trojan virus detection, analyse the Trojan packet and store its information for further viewing using any web browser.

The methodology includes three main parts, 1st part Ubuntu (Operating System) , 2nd part software design using packet capture, and the last part analysis packets applied under the operating systems Ubuntu and Windows. Ubuntu is a computer operating system based on the Debian Linux distribution; Ubuntu provides an up to date, stable operating system for the average user, with a strong focus on usability and ease of installation.

Pcap (packet capture) consists of an Application Programming Interface (API) for capturing network traffic. Unix-like systems implement pcap in the libpcap library; Windows uses a port of libpcap known as WinPcap. Monitoring software may use libpcap and/or WinPcap to capture packets travelling over a network and, in newer versions, to transmit packets on a network at the link layer, as well as obtain a list of network interfaces for possible use with libpcap, ALO support saving captured packets to a file, and reading files containing saved packets; applications can be written, using libpcap to be able to capture network traffic and analyze it, or to read a saved capture and analyze it, using the same analysis code. A capture file saved in the format that libpcap and use can be read by applications that understand that format.

## Software Design

Software design is a multi-disciplinary activity that develops tools through effective communication of ideas and the use of engineering practices. The process is passing through at five phases as below:

Phase 1: Capture and extract network packet information.

Phase 2: Check Trojan infected packet.

Phase 3: Analysis of Trojan packet.

Phase 4: Store Trojan packet information in a file

Phase 5: Display information using web browser.

Phase 1: Capture and extract network packet information.

Packet Sniffer is used to capture network packets information and stored into a data buffer for further processing. The Packet Sniffer module operates at the network layer and captures network packets physically across through the Network Interface Card (NIC). In this module the NIC receives packets directly from a network segment. The process of this module involves network hub setting, packet capture and packet information extraction and packet information storage into a file. The functionality of the probe module is realized through the usage of libpcap open source library [4].All processes in this module utilized *libpcap* library functions, as shown in Figure 1.

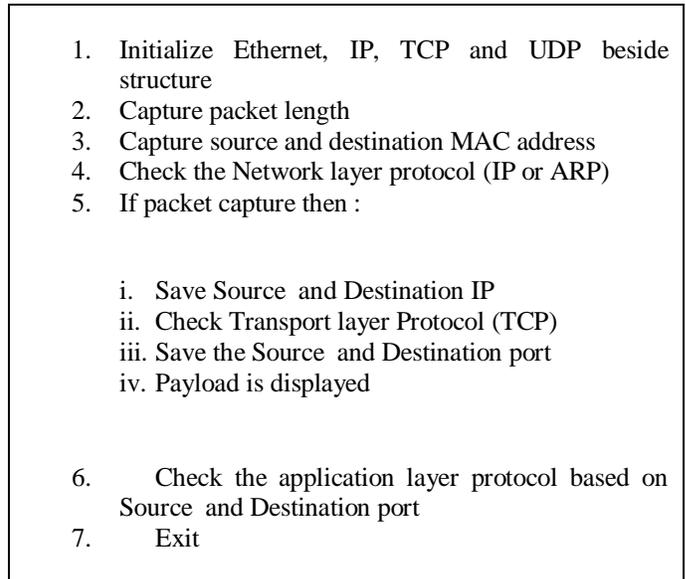

1. Initialize Ethernet, IP, TCP and UDP beside structure
2. Capture packet length
3. Capture source and destination MAC address
4. Check the Network layer protocol (IP or ARP)
5. If packet capture then :

    i. Save Source and Destination IP
    ii. Check Transport layer Protocol (TCP)
    iii. Save the Source and Destination port
    iv. Payload is displayed

6. Check the application layer protocol based on Source and Destination port
7. Exit

Figure 1: The Algorithm Process for Packet Information Grabbing

*Phase 2: Check Trojan infected packet.*

Packets that have been captured through the network segment are displayed in Figure 2. The packets crossing the network are scrambled or not in readable mode, packets that pass through are in binary form therefore Packet Sniffer is designed to print data in hex and ASCII format. The data are printed in rows of 16 bytes and the payload number (in bytes) is defined in line 2. This type of definition is easier to use in detecting normal packet and infected packets.

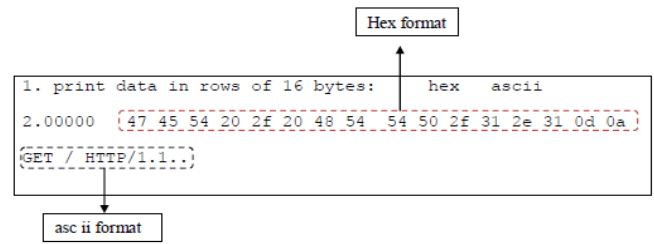

Figure 2: Packets captured through the network segment are displayed

Phase 3: Analysis of Trojan packet.

The analyses of a Trojan packet in the previous method entailed comparing two normal and abnormal packets. After the abnormal packet had been is detected, the packets were analyzed to determine whether they are Trojan packets or not. Two types of Trojan (Trojan horse and Backdoor) were ought, and both types were analyzed in the same way the detection is explained on the TCP header.

The analysis involved four steps as follows:

    **Step 1**: Analyzing Ethernet frame,
    **Step 2**: Analyzing Internet Protocol,
    **Step 3**: Analyzing TCP protocol, and
    **Step 4**: Analyzing the payload pattern.

Phase 4: Store Trojan packet information in a file.

All the information obtained from the infected packets was collected and stored into a valid file the web browser as shown in Figure 3.

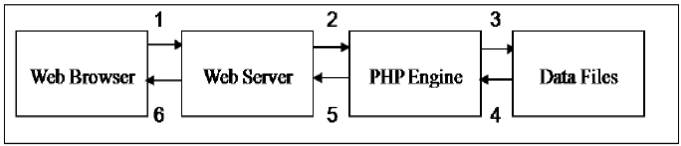

1) If the packets are infected by Trojan attack then:

    i.  Open a file in append mode for storing packet info.

        • Packet grabbing data and time

        • Packet type (TCP/UDP)

        • Packet source and destination MAC address

        • Packet source and destination IP

    ii.  Close file

Figure 3: Infected Packet Storage Module

Phase 5: Display information using web browser.

Packet Sniffer can capture and extract all the packet information that has been defined, then it will analyze all infected captured packet information and store all necessary required information into files for viewing. The possible information includes the protocols being used on a network segment, but it concerns mainly the behavior of network TCP header protocol, IP header protocol and traffic between each source and destination. The analyzed network traffic information can be viewed through a web browser the data transaction between web browser and Packet Sniffer consists of some sequences of actions, which are shown in Figure 4.

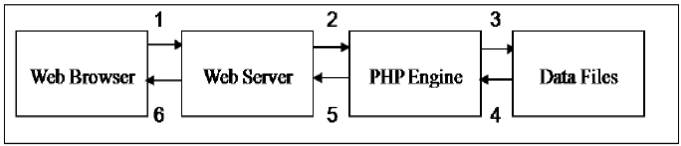

Figure 4: Data Transaction between Web Browser and Server

A user's web browser issues an HTTP request for a particular web page. The web server receives the request for results. PHP script retrieves the file and sends it to the PHP Engine for processing. The PHP engine will finish running the script, which usually involves formatting the results in HTML. It then returns the resulting HTML to the Web Server. The Web Server passes the HTML back to the browser, where the user can view the requested output. The web pages are developed using PHP and HTML code. The analyzer analyzes the packet information and stores it in data log files. The PHP script reads the files according to user selection and displays the internet traffic information. For ease and efficiency, a web-based user interface is used. A web-based interface eliminates problems of porting, while a single script provides uniform results, regardless of the operating system, wherever the user is located.

## VII. RESULTS

The results of the implementation to the sniffer packet network for the operating systems Linux and Windows are as follows:

### A. Results for Trojan horse attack payload

The attack payload is obtained by sending an e-mail to the PC an (.exe) file named hp- ftp is attached in the mail and sent to the PC and the file may be downloaded. The attached file which contains the Trojan horse has the following behaviors:

1) Contains net stat information to abort [at] yahoo.com LinuxPir8 [at] yahoo.com, see Figure 5.

2) File size:14140 : the infected packets, i.e the Trojan horse that was tested for this experiment has certain information, such as length or file size and net stat information. Figure 6 clearly shows the states from the TCP segment byte 0230-12e0 the payloads are not infected However, when the net stat information is encrypted or the file is in process, the payloads after that are infected. The same thing occurs to the information available for the Trojan horse (file name hp-ftp) where file size is 14140, when the network protocols processes the file information it becomes infected, as shown in Figure 6.

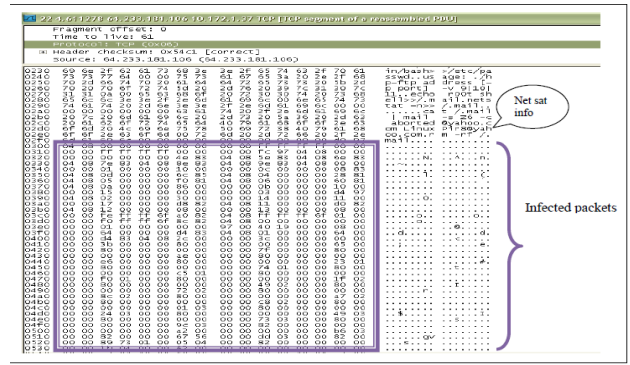

Figure 5: Sniffing Results for HP-FTP file –Net sat inf.

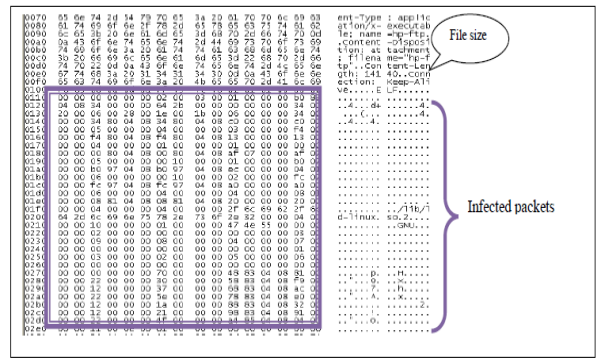

Figure 6: Sniffing Results for HP-FTP file – file size.

3) To establish whether or not this file has been infected by a Trojan horse, the file is placed independently in specified folder. After that Anti-virus program is used to scan the specified folder and examine whether the program is malicious or not.

*B. Results for Backdoor Trojan attack payload*

An attack payload is obtained by sending an e-mail to a PC an (.exe) file named "backdoor" is attached in the mail and sent to a PC the file is then download. The attached file which contains Backdoor Trojan has the following behavior.

- The results for the Trojan Backdoor attack payload for Windows based Wireshark, show that the Trojan backdoor produce a different pattern of behavior compared with Trojan horse of normal behavior. The Trojan Backdoor output is empty payload. No data are "backdoor"based on the results the empty payload is defined as an attack, compared with normal packets. Figure 7 shows the control bit behavior.

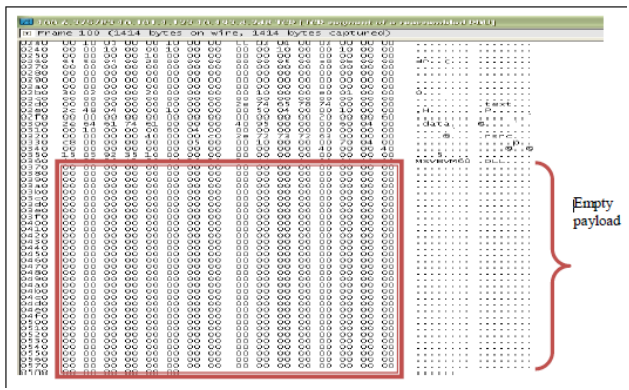

Figure 7: Sniffing result for backdoor.

- The flags analysis for backdoor is based on Linux and Windows. The discussion is based on a comparison of the control bits.

    In the Windows operating system the flag captured at time 10.438 shows the normal behavior of control bits. Flag ACK is set at the time with sequence time 79225 the acknowledgment number is 759 the TCP flow started off well, without any abnormalities. With the same attempts in Ubuntu , at time 16.479, 16.480 and 16.480 three- way handshake occurs showing that a TCP connection is established. Abnormalities occur in windows at time 10.712 only the SYN flag is set which means it has initiated a TCP connection, as shown in Figure 8.

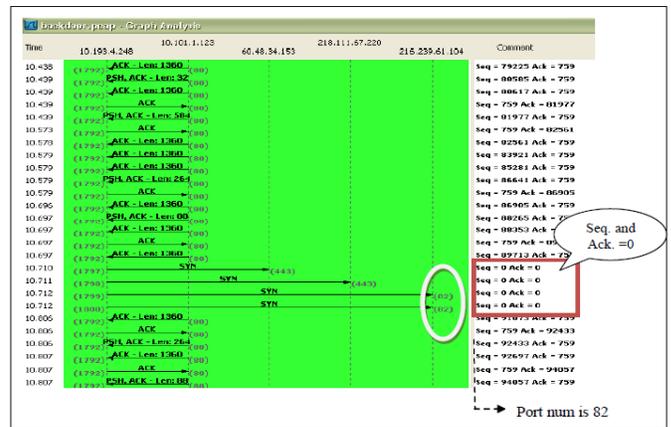

Figure 8: Analysis Graph for Backdoors based Windows

- The port number changed from 80 to 82, which shows a weird abnormal behavior. TCP by right should only access to port 80 since there was no other network access in the Internet compared with Linux based Wireshark all the destination port numbers were 80. From these abnormal behaviors it can be concluded that Backdoor Trojan infects port numbers, also at time 10.710 till 10.712 the sequence number and acknowledged number was 0.

    - Empty data are transmitted through the TCP flow. According to this behavior, it can be concluded or there can be strong agreement that this is an abnormal behavior of TCP flow caused by a malicious code (Backdoor Trojan).

    Table 1 shows the behavior of Trojan horse and Backdoor in both Linux and Windows.

Table 1: Comparison Trojan virus between Linux and Windows

| Events | Linux | Windows |
|---|---|---|
| Operating System behavior | GNU/Linux is more stable than windows. | Windows is easily attacked without user's knowledge. |
| Trojan horse behavior | The packets are infected after the information enters the network. | The packets are infected after the information enters the network. |
| Backdoor Trojan behavior. | Flags analysis for backdoor based Linux. The discussion is based on comparison of both figures based on the control bits. The flag captured at time 10.438 shows the normal behavior of control bits. | Flag ACK is set at the time with sequence time 79225 and acknowledgment number is 759, TCP flow stated of well without any abnormalities |

| The Trojan horse attack. Packets infected Sniffer - Payload. | The packets are not fully infected they are just infected in certain parts. | Both Wireshark and Packet Sniffer have the same results for Trojan horse attack. |
|---|---|---|
| Execution the Trojan Backdoor (.exe). | In GNU/Linux (Ubuntu), The file unable to run and shows an error message. | File can only execute in Windows. |

## VIII. DISCUSSIONS

*A.* Trojan Backdoor which is in (.exe) is executed in both GNU/Linux (Ubuntu) and Windows operating systems. In GNU/Linux (Ubuntu), on should double click the download file and try to run the file. The file cannot run and shows an error message, where the programs are in Windows can be executed. Upon execution of the backdoor .exe file, the file will enter the system and might be able to crash the system. The file has valid icon which can be executed.

*B.* Analysis of infected packet using Linux based Wireshark vs. Packet Sniffer: the results that were obtained from Trojan horse show that it is attack payload for Linux-based Wireshark and Packets Sniffer. Both Wireshark and Packet Sniffer have the same results for a Trojan horse attack. Windows-based and Linux-based Wireshark also give the same output. Therefore the output for Linux-based Wireshark and Packet Sniffer is discussed together. A Trojan horse change the behaviors of packets are infected after the information enters the network. The packets are not fully infected, only certain parts are infected, 00 means that no packet has been sent to the destination port from the source port. In this case, when downloading the file TCP data transfer was interrupted by comparison with the normal packet, the (.exe) file did not interrupt TCP connection process. The only difference is that the executable file can be executed in Windows only and Linux does not allow the execution.

*C.* Analysis of infected packet in Windows-based Wireshark the results will discuss the Trojan Backdoor attack payload for Windows-based Wireshark .To access the web page, first we have to type in the link as stated the local host is used since in have created my own server and have to insert the host PC's IP address for access from another PC.

## IX. CONCLUSION

The main target of this paper is to detect Trojan horse infected packets from a computer network segment before they can attack a computer and compare the attacked "Trojan horse and backdoor" through Ubunto (Linux) and Windows. From the implementation we can conclude that:

- Linux and Windows have the same output for a Trojan horse attack through the infected packet based Wireshark vs. output.

- The infected captured packet for both Linux and Windows have the same behavior. Besides that, the comparison between normal packets and Trojan packets shows that there are differences between the payload which is found only inside the packet payload.

- The objectives of this paper have been partially achieved in the following sense first infected packet based Wireshark vs. output has been captured and network information has explored and secondly Trojan attack from a computer network segment has been detected and monitored.

- The Packet Sniffer that uses Linux command has successfully captured live data, and this tool allows the sniffing of more packets compared with Wireshark, also Wireshark sniffs packets very fast, compared with Packet Sniffer which allows the user to capture up to 1000 packets.

- The designed code is able to capture TCP, IP, UDP and also ICMP protocol information. The TCP payload was used to obtain more in depth information of packets to detect a Trojan attack.

## REFERENCES

[1] Antivirus Scanner for Unices. cited; Available from: http://www.bitdefender.com/world/business/antivirus-for-unices.html

[2] Bishop, M., "An Overview of Computer Viruses in a Paper Environment", p. 1-32, Technical Report: PCS-TR91-156-1999.

[3] Danchev, D. ,"The Complete Windows Trojans", cited; Available from: http://www.windowsecurity.com/whitepapers/The_Complete_Windows_Trojans_Paper.html. Aug 29, 2005.

[4] Jacobson, V., Leres, C., & McCane, S. Tcpdump Manual Page. Retrieved January 10, 2008, from http://www.tcpdump.org ,1997.

[5] Li, X., Computer Viruses: The Threat Today and The Expected Future. p. 71.,2003.

[6] P2P-Worm.Win32.BlackControl.g, Trojan Programs. [cited; Available from: http://www.securelist.com/en/descriptions/15243378/P2PWorm.Win32.BlackControl.g, Aug 20, 2010.

[7] Shimonski R J, "Trojan Horse Primer", http://windowsecurity.com/articles/Trojan_horse_primer.html ,2004.

[8] Trojan Programs. cited; Available from: http://www.viruslist.com/en/virusesdescribed?chapter=152540521, Oct 20,2010.

[9] Ubuntu Operating System. cited; Available from: http://en.wikipedia.org/wiki/Ubuntu_(operating_system)

[10] Wireshark. cited; Available from: http://en.wikipedia.org/wiki/Wireshark , November 2010.

AUTHORS PROFILE

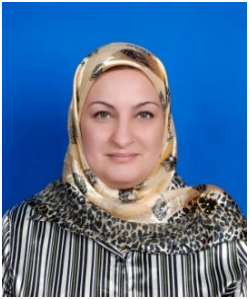

Ghossoon M.W. AlSaadoon is senior lecturer in Network security & DataBase, Dr.Al-Saadoon is a Head Department of Management Information Systems & Director of Academic Staff Performance Development Center at the Applied Science University College of Administrative Science, Manama, and Kingdom of Bahrain. She holds PhD degree in computer science from the Iraqi commission for computers & informatics /institute for post graduate studies in informatics, 2006 in addition; she is a membership of CSC-journals & ISACA member. Dr. Al-Saadoon has more than 19 years of experience including project management experience in planning and leading a range of IT-related projects. Dr. Al-Saadoon supervised many computer and communication engineering students leading to Ph.D. and M.Sc. degree in computer and communication engineering in UniMap.

Dr. Al-Saadoon has three awards (two from Ministry of Science Technology and Innovation (MOSTI) and One from UniMap University).

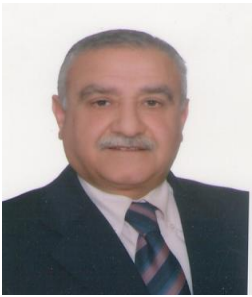

Hilal Mohammed Yousif Al-Bayatti is university vice president and professor of computer science with applied science university, kingdom of Bahrain, where he taught and conducted research in computer and information security.
He earned his Ph.D. in computer science at Loughborough University of Technology (U.K), his M.Sc. in computer science at University College London (U.K) his B.Sc. in mathematics at Baghdad University (Iraq). Prof. Hilal has been teaching for 25 years undergraduate and graduate students in computer science fields. Prof Al-Bayatti supervised many computer science students leading to Ph.D. and M.Sc. degree in computer science.

Prof. Hilal has published more than 50 referred research papers in leading journals. He is member of many steering and technical committees of national and international conferences.